\documentclass[conference]{IEEEtran}
\IEEEoverridecommandlockouts
\usepackage{cite}
\usepackage[english]{babel}

\usepackage[T2A,LGR,T1]{fontenc}
\usepackage[latin9]{inputenc}
\usepackage{array}
\usepackage{float}
\usepackage{textcomp}
\usepackage{multirow}
\usepackage{stackrel}
\usepackage{graphicx}
\usepackage{setspace}
\usepackage[export]{adjustbox}
\usepackage{algorithm}
\usepackage{caption}
\usepackage{subcaption}
\usepackage{algpseudocode}
\usepackage{enumerate}
\usepackage{xcolor}
\usepackage{graphicx}
\usepackage{amssymb}
\usepackage{color}

\usepackage{amsmath,amssymb,amsfonts}
\def\BibTeX{{\rm B\kern-.05em{\sc i\kern-.025em b}\kern-.08em
    T\kern-.1667em\lower.7ex\hbox{E}\kern-.125emX}}

\begin{document}

\title{FLCC: Efficient Distributed Federated Learning on IoMT over CSMA/CA} 
\author{
  Abdelaziz Salama\textsuperscript{$\star$}\qquad Syed Ali Zaidi\textsuperscript{$\star$} \qquad Des McLernon\textsuperscript{$\star$} \qquad Mohammed M. H. Qazzaz\textsuperscript{$\star$}\\ 
  \textsuperscript{$\star$}School of Electrical and Electronic Engineering, University of Leeds, Leeds, UK\\
  \textsuperscript{$\star$}Corresponding author: Abdelaziz Salama (elamasa@leeds.ac.uk)
}


\maketitle

\begin{abstract}

Federated Learning (FL) has emerged as a promising approach for privacy preservation, allowing sharing of the model parameters between users and the cloud server rather than the raw local data. FL approaches have been adopted as a cornerstone of distributed machine learning (ML) to solve several complex use cases. FL presents an interesting interplay between communication and ML performance when implemented over distributed wireless nodes. Both the dynamics of networking and learning play an important role. In this article, we investigate the performance of FL on an application that might be used to improve a remote healthcare system over ad hoc networks which employ CSMA/CA to schedule its transmissions. Our FL over CSMA/CA (FLCC) model is designed to  eliminate untrusted devices and harness frequency reuse and spatial clustering techniques to improve the throughput required for coordinating a distributed implementation of FL in the wireless network. 

In our proposed model, frequency allocation is performed on the basis of spatial clustering performed using virtual cells. Each cell assigns a FL server and dedicated carrier frequencies to exchange the updated model's parameters within the cell. We present two metrics to evaluate the network performance: 1) probability of successful transmission while minimizing the interference, and 2) performance of distributed FL model in terms of accuracy and loss while considering the networking dynamics. 

We benchmark the proposed approach using a well-known MNIST dataset for performance evaluation. We demonstrate that the proposed approach outperforms the baseline FL algorithms in terms of explicitly defining the chosen users' criteria and achieving high accuracy in a robust network.
\end{abstract}

\begin{IEEEkeywords}
Federated Learning, CSMA/CA, IoT Privacy.
\end{IEEEkeywords}

\section{Introduction}

Many more gadgets and edge devices are online nowadays, including smartwatches, house assistants, security systems, and even remote healthcare to improve the quality of life for people such as the elderly and individual patients in their houses. The data generated by these devices has become important for driving innovative solutions through Machine Learning (ML) applications. Several applications, such as autonomous vehicles and healthcare systems, are rapidly deploying IoT solutions \cite{li2013towards}. However, because the data is often large in quantity and contains private information, applicants will often will be prevented from logging into a data centre. Consequently, any collaborative IoT solutions without proper measures for privacy preservation will suffer low uptake. Therefore, Google launched Federated Learning (FL) in 2017 \cite{mcmahan2017communication} as a promising approach to train AI models on personal data dispersed over billions of devices while mitigating privacy leakage risks and improving collaboration between users.

\subsection{Motivation}
FL has proven to be an effective solution for reducing the workload on the server and the amount of data that needs to be transmitted. Moreover, FL addresses the privacy problem and provides capabilities to train ML models on local data, which can then be shared globally to accelerate learning processes. The shared models are often combined by a central server, and aggregated models are shared back with the end devices. The performance of FL algorithms is also dictated by the communication between the participating devices and the centralized server. 

Now when FL is implemented on wireless nodes, propagation, topology and medium access dynamics all dictate the FL performance. While centralized FL can be implemented easily in cellular IoT deployments, ad-hoc wireless networks warrant distributed FL algorithms. In distributed FL implementation, models are just shared in a defined neighbourhood so that the learning of the desired ML task can be accelerated through collaborations locally. Such a model presents an additional challenge as the simultaneous uplink transmissions from devices may generate additional interference or collision at the application layer. These will in turn increase the convergence time for ML algorithms. 

In order to address the above-mentioned challenge, a co-design of communication medium access control (MAC) protocols (e.g., CSMA/CA) with the distributed FL task must be considered. For such a system, an optimal configuration of design space could yield significant performance gains. CSMA/CA is the primary MAC protocol employed by 802.11 network deployments \cite{ieee2007part}. However, in such networks, the FL model transmission may suffer from interference when servers are within the transmission range of different devices utilising the channel simultaneously which can negatively impact the learning process in real-environment applications.
 
This paper proposes a distributed FL model to improve IoT solutions applied to healthcare, which is most commonly known as the Internet of Medical Things (IoMT) \cite{vishnu2020internet}. The FL model can perform training locally on different IoMT devices (i.e., mobile, wearable devices and MRI devices) where a server collects only the traffic model parameters from these massively distributed IoMT devices in a large-scale network. These IoMT devices could be implemented to monitor people, such as diabetes patients and the elderly and then advise the recommended nutrition and care according to an intelligent collaborative FL model, which can lead to better human health outcomes. FL can provide the following significant benefits for IoMT applications in various medical areas according to its innovative operational concept.
\subsubsection{Preserving Privacy} Building resilient and secure IoMT systems requires the ability to protect user information, especially in the context of the increasingly strict data privacy protection laws. The FL system addresses the data privacy leakage issue because only the local updates are needed by the central server for cooperative ML training, while the local data never leaves the device. 
\subsubsection{ Minimize Latency} FL can considerably lower communication costs in intelligent IoMT networks, such as latency. These IoMT networks are constrained by huge amounts of raw data transfer, and by using FL we can avoid offloading large volumes of data to a server. As a result, FL contributes to the reduction of network spectrum resources needed for data training. 
\subsubsection{ Improved Learning Quality} FL benefits from significant dataset resources from many IoMT devices over a large-scale IoMT network to collaborate in solving an ML problem. This collaboration can speed up the overall training process and convergence rate and improve learning accuracy that centralized traditional ML approaches might not be able to achieve.
\subsection{Key Contribution}
Our proposal considers the optimal parametric configuration for CSMA/CA protocols to eliminate the collision and minimize the interference rate while considering the implementation of distributed FL for IoMT and healthcare systems. We demonstrate that by optimally dimensioning the systems through spatial clustering and frequency allocation mechanisms, where distributed FL mechanisms can outperform baseline approaches and precisely evaluate the wireless communication in the network during the learning process. Overall, the core contribution of this paper is to present a study of distributed FL over 802.11 networks for the IoMT and healthcare systems. 
\subsection{Paper Organization}
The rest of this paper is organised as follows. In Section \ref{sec: System model}, we introduce the FL based on CSMA/CA MAC and algorithm architecture system. Section \ref{sec: Simulation and results} includes the simulation environments and criteria used to simulate the FL over CSMA/CA (FLCC) model. We then evaluate the outputs metrics in comparison with an FL model over a traditional CSMA/CA scheme (i.e., baseline models) applying different network intensities. And finally,  conclusions are presented in Section \ref{sec: conclusion}.
\section{System model} \label{sec: System model}
\subsection{Topological Considerations}
We consider that wireless nodes (i.e., IoMT devices) are randomly scattered in a 2-D plane. Their location can be modelled as a homogeneous Poisson point process (PPP). 

\begin{figure}
\includegraphics[scale=0.26]{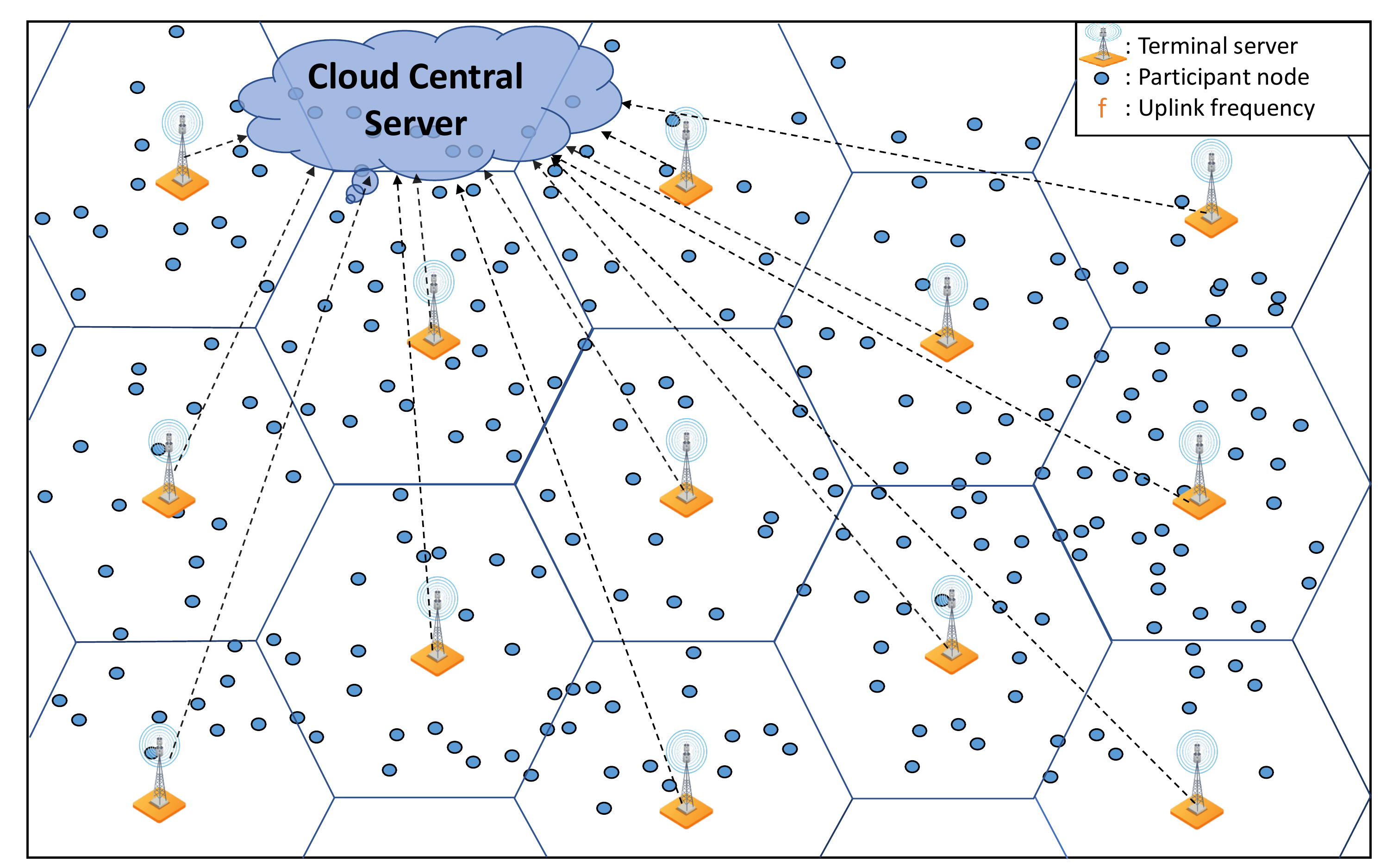}
\caption{Node distribution and hexagonal cell layout for the FLCC network.}\label{fig:my_label}
\end{figure}

The density of the PPP is assumed to be $\lambda$ as shown in Figure \ref{fig:my_label}. We assume these devices share the model's parameters with a central FL server. For analytical tractability, we assume that the 2-D plane is tessellated using hexagonal lattices. Each hexagonal lattice is assigned an FL central server, which is placed in the centre and allocated its own frequency range as shown in Figure \ref{fig:my_label}. 

The key motivation behind clustering the network, dividing the area into subareas (cells), and adopting a frequency reuse approach is to minimize interference \cite{DIRANI2011141}. We assume that the network has an FL server in the central of each cell and the nodes (i.e., IoMT participants) distribute in a PPP\cite{haenggi2009stochastic} of the server of intensity ($\lambda$) over uniform areas as shown in Figure \ref{fig:my_label}. The transmitter's nodes adopt the Request to Send/Clear to Send CSMA/CA approach where the node can only share parameters with the server when no one of the neighbours within the same cell is transmitting. Otherwise, it must wait and attempt to transmit again in a random backoff-time. Furthermore, the server is assumed to have a multi-input multi-out antenna. Hence it can receive parameters from different participants simultaneously up to the maximum number of sub-channels in the designed network and broadcast the global model update on predefined channels. 

In this work, we consider a typical receiver that is connected to the desired transmitter. For the small-scale path-loss model, a Rayleigh fading channel is chosen and complemented with a single slope large-scale path-loss. Thus, the received signal from the desired transmitter at the FL server (i.e., typical receiver) is ($P_{k}h_{ko}d_{ko}^{-\alpha}$) \cite{jing2012achievable}, where $P_{k}$ is the power for the transmitted signal from the desired transmitter device $k$; ($\alpha\ge2$) is the path-loss exponent and finally, $d_{k0}$ and $h_{ko}$ denote the distance and the fading coefficient for the channel between device $k$ to the target FL server, respectively. 

In order to examine the real environment scenario in terms of interference within the network and explicitly define the number of IoMT devices that successfully transmitted, we implement a physical interference model using the well-known Signal-to-Interference-and-Noise Ratio (SINR) and express the successful transmission probability ($ P_{s}$) of parameters for the transmitter ${k}$ in the FL training process as follows:
\begin{align}
P_{s}= P(SINR\ge T) & =P\left(\frac{P_{k}h_{k0}d_{k0}^{-\alpha}}{\sum_{i\in\varphi}I_{i}+N_{0}}\ge T_{k}\right)\label{eq: SINR prob of succc}
\end{align}
where $T$ is a predefined threshold, $I_{i}$ is the interference from the device $i$ in the network (i.e., $I_{i}=P_{i}h_{i0}d_{i0}^{-\alpha}a_{i}$), with $i=1,2,..., A$; $A$ is the total number of active IoMT devices within the desired coverage area and $i\in\varphi$ where $\varphi$ shows the number of all IoMT devices within the whole target area. Now $b_{i}$ is a binary random variable that defines the state of the device, whether in transmitting mode or ready to receive, such that $P_{A}=\,$Probability${\{b_{i}=1\}}$ shows the IoMT device is in transmitting mode and $1-P_{A}=$ Probability${\{b_{i}=0\}}$ represents the device is ready to receive and is not transmitting.

For an FL network over CSMA/CA simulation, we obviously need models that define not only whether a participant's parameters reach a server once the channel is available (i.e., carrier sense indicates no traffic on the channel), but also whether the quality of service and proper capacity are satisfied or not. For that matter, maximum network capacity $(C_{CSMA})$ and SINR \cite{gupta2000capacity} with predefined threshold ($T_{k})$ condition needs to be addressed as follows:
\begin{align}
C_{CSMA} & =\text{log}_{2}(1+T_k)P(SINR\ge T_{k})\label{eq: Max Cap CSMA}
\end{align}
where $P(SINR\ge T_{k})$ is the probability that SINR at a typical receiver (i.e., FL central server) is higher than a certain threshold $T_{k}$.. In our proposed FL network, $T_{k}$ is assumed to be a variable that affects the maximum transmission capacity that the network can achieve. The derivation of maximum transmission capacity per cell in bits/s/Hz/participant has been addressed in \cite{jing2012achievable}.

~~From (\ref{eq: Max Cap CSMA}), the intuition is obtained as to what extent the parameters of network $T_{k}$ and SINR affect the maximum achievable transmission capacity of the FL based on the CSMA/CA network, where a trade-off between the probability of successful transmission and the SINR threshold is required.

The centralized FL approach is designed to work in a centralized environment where there is a centralized server handle to the aggregated models' updates in the network so as to build an efficient global model. Consequently, the FL network needs to be designed with respect to minimizing interference which is an important task that leads to having a reliable and efficient FL algorithm performance. 

\subsection{Performance Metrics}
There are several performance metrics that can be used to evaluate the performance of an FL system. In this work, we use the two most common performance metrics in ML to evaluate our classification FL model which are as follows:

(a) Categorical cross-entropy loss which is a loss function commonly used for classification tasks with more than two classes. It is often used in conjunction with softmax activation in the output layer of a neural network \cite{ho2019real} 

(b) Accuracy which is a metric commonly used to evaluate the performance of an ML model. It is the fraction of correctly classified instances (true positives and true negatives) out of all instances in the dataset. In federated learning, accuracy can be used to improve the performance of the system by helping to identify and address any issues with the model's predictions. To calculate accuracy, the model's predictions are compared to the true labels for a given dataset. The number of correct predictions is then divided by the total number of instances in the dataset to produce the accuracy score.

\subsection{FL Network and Algorithm Formulation} \label{sub: Fl algor}
The principle concept of the FL learning process is a distributed ML approach that preserves the privacy of the terminal nodes where the local data never leave the terminal and shares only the local model parameters with others through a central server. The designed network considers a set of individual nodes ($\varphi$ ) in which $\varphi={1,\ldots,K}$ is randomly distributed as a spatial point process following a stationary PPP \cite{haenggi2009stochastic}. The server is allocated in a certain central cell surrounded by a group of cells distributed uniformly where each neighbour's cell to the central cell assign a dedicated frequency to exploit by the users within the spatial location of the cell for sharing the local models as shown in Figure \ref{fig:my_label}. 

The communication of multiple participants with a centralized server is controlled by the CSMA/CA MAC protocol. In the FLCC network, we propose having a number of uplink channels ($N$) and a uniform distribution of cells over a particular area subject to the condition that the sensing range of each participant reaches the other participants within the same cell. This will play an essential role in minimising the collision from hidden nodes and then have an improvement in the training process. Consequently, $N$ participants from different $N$ cells (i.e., the server's cell and the surrounding cells) can share their parameters simultaneously using the $N$ dedicated uplink carrier frequencies, and the server can broadcast the global model update on a certain downlink channel to the participants within the network. 

At the steady state, each participant trains his local model using the local dataset and then has a matrix of updated parameters to send to the server at a random time slot. The central server aggregates the available local models to build a global model. In this work, the aggregated parameters are not treated equally as the network segmentation will divide the users within the network into subgroups. Each user ($i$) has a weighted probability ($p_i$) between 0 to 1, where 0 represents the unwanted devices that will be eliminated from updating the global model (e.g., discovered hackers and militia devices) at this iteration, and closest to 1 represents the most efficient devices for the learning process in the network (e.g., well-trusted devices) whereas the summation of the devices' weighted probabilities at each iteration must be 1 (i.e., $\sum_{i}{p_i}=1$). Nevertheless, the device's weight probability within the network will be evaluated regularly via the server using rewards and penalties in reinforcement learning approaches (e.g., Q-learning algorithm) to update their assigned weighted probability based on their updated behaviour. In other words, every action the IoMT participants take in the learning process is recorded as an "action," and each action advances the IoMT participant to the next state. This operation is noted and rated as a "reward" if it improves the learning metrics or a "penalty" if it is not \cite{sutton2018reinforcement}. 

Afterwards, each server aggregates the parameters and averages them with the cloud central server model (i.e., the latest cloud central server model result from averaging the aggregated global models from the cooperative servers in the previous iteration) to update the global model at each server within the large-scale network and then broadcasts it to both the cloud central server and the participants to improve the local models' performance for the next iteration. This process continues updating for several sequence iterations according to cooperation between the participants, each cell server and the cloud central server until achieving the predefined convergence condition in the network.

\subsection{FL over CSMA/CA learning criteria} \label{sub: Fl algor implem}
In the learning process, each node $i\in\varphi$ has its local dataset $\mathfrak{O}_{i}$ which is heterogeneous or homogeneous datasets that follow an unknown probability distribution $p(x,y)$ and possibly has non-empty interaction for different samples datasets (i.e., $\mathfrak{O}_{i}$ and~~$\mathfrak{O}_{j}$~~where $i\neq j$). The samples consist of $n$ instance-label pairs of samples$(X_{i}^{n},Y_{i}^{n})$, where $X_{i}^{n}$ and $Y_{i}^{n}$ represent the labelled sample input and output respectively for a specific learning objective in the device $i$, and $n=1,....,k$ where $k$ is the total length of $i$ participant's dataset. Each instance $X_{i}^{n}\in X_{i}\subseteq X$, where $X_{i}$ denotes the local instance space of participant $i$ and $X$ denotes a global instance space, which satisfies $X\subseteq\bigcup_{i=1}^{\varphi}X_{i}$ (i.e., the total datasets for all devices in the network). 

Consequently, there are ${X_{i}^{(1)},X_{i}^{(2)},.....,X_{i}^{(k)}}$ i.i.d samples at each participant node in the proposed network, and the total number of examples $k$ is a variable depending on the size of the local datasets for each one.

All participant devices have to have the same machine-learning model (e.g., a CNN), which has a common weights parameters matrix ($\textbf{W}$). One of the main aims of the designed model is to minimize the mean cross-entropy loss between the expected (actual) output and the predicted output for all participants: 
\begin{align}
\stackrel[\textbf{W}]{}{\text{arg min}}F(\textbf{W}) & \triangleq\frac{1}{A}\sum_{(i\in\varphi)}f_{i}(\textbf{W})\label{eq: min loss fun 20}
\end{align}
where $F(\textbf{W})$ denotes the global loss function, $f_{i}(\textbf{W})$ represents
the local loss for the device $i$ and $A$ represents the length of
active devices that succeed to transmit to the server at each iteration in the training process where: 
\begin{align}
i & =1,2,\ldots..,A,\forall\,\,\,SINR\thinspace\,\ge\thinspace\,T_{k}. \label{eq: succ set}
\end{align}

During the learning process, each participant $i$ finds the local model loss function based on the local training dataset as follows:
\begin{align}
f_{i}(\textbf{W}) & =\frac{1}{M}\sum_{m=1}^{M}l\biggl(h_{\textbf{W}}(X_{i}^{m}),y_{i}^{m}\biggr),\label{eq: cross-entropy 21}\\
 & i\in\varphi\nonumber
\end{align}
where $M=\frac{|\mathfrak{O}_{(i)}|}{B_{i}} $, $y_{i}^{m}$ is the actual output and $h_{W}(X_{i}^{m})$ is the hypotheses output of the model based on the parameters matrix ($\textbf{W}$) for instance inputs $X_{(i)}$ of $m$ subset of the sample training $\mathfrak{O}_{i}$ for the participant $i$ and $B_{i}$ is a hyperparameters batch size, and $l(.)$ denotes the loss in the model at each sample (e.g., categorical cross-entropy loss function \cite{ho2019real}). 

Afterwards, each participant $i$ finds the gradient matrix $\nabla f_{i}(\textbf{W}_{i}^{t})$ based on local training to update the local models' weights via a local optimizer such as stochastic gradient descent (SGD) which is shown as follows:

\begin{align}
{\nabla f}_{i}(\textbf{W}_{i}^{t})=\frac{1}{M}\sum_{m=1}^{M}\nabla l\biggl(h_{(\textbf{W}_{i}^{t})}(X_{i}^{m}),y_{i}^{m}\biggr),\label{eq:22} \\
\textbf{W}_{i}^{t}:=\textbf{W}_{i}^{t}-\eta_{i}\nabla f_{i}(\textbf{W}_{i}^{t}).\,\,\,\,\,\,\,\,\,\,\,\,\,\,\,\,\,\,\,\,\,\,\label{eq: 23}
\end{align}

Hence, $\nabla f_{i}(\textbf{W}_{i}^{t})$ is obtained at each iteration $t$ from the $i^{th}$ participant and updates the weights matrix using the predefined hyperparameters of SGD (i.e., learning rate ($\eta$) and batch size $B_i$) on the participant local data after a number of local training. Subsequently, the gradient $\nabla f_{i}(\textbf{W}_{i}^{t})$ in the learning represents the changing rate of $f_{i}$ with respect to the model parameters $\textbf{W}_i$ at iteration $t$, and the same update rule is applied to every participant in the network simultaneously. Afterwards, the global model is processed by aggregating the local models from the participants. In our design, the server will be able to receive multiple model parameter updates from many participants simultaneously as described in the subsection \ref{sub: Fl algor}. The total number of participants ($A$) at each iteration is defined as a summation of devices that satisfy the communication constraints and their SINR is higher than the predefined threshold ($T$). 

For a global optimization and collaboration between IoMT devices within the network, our global model ($\hat{\textbf{W}}$) will be updated using the Federated Averaging algorithm (FedAvg) \cite{mcmahan2017communication} which is efficient and simple to apply, and in our proposal we consider each participant' weighted probability $p_i$ in the learning process at each iteration, and thus the global model update can be summarized as follows:

\begin{align}
\hat{\textbf{W}}^{t} & =\biggl(\sum_{i=1}^{A}p_{i}*{\textbf{W}}_{i}^{t}\biggr)\label{eq: eq 24}
\end{align}
\begin{align}
\hat{\textbf{W}}^{t} & =\textbf{W}^{(t+1)}.\label{eq: eq26 new para}
\end{align}

At the next iteration, the participants receive the new global model to update their local models and validate the performance on the local data and then the server will decide whether the network achieves the predefined convergence condition ($\epsilon$) or less to end the learning process or not by averaging the latest aggregated gradient ($\hat{\nabla}F(\hat{\textbf{W}}^{t})$) from the participants within the network as follows: 
\begin{align}
\,\,\,\,\,\,\,\,\,\hat{\nabla}F(\hat{\textbf{W}}^{t}) & =\frac{1}{A}\sum_{i=1}^{A}{\nabla f}_{i}(\textbf{W}_{i}^{t})\, \leq \,\epsilon \label{eq: eq25}
\end{align}
 
 The summary of the implemented algorithm is given in Algorithm (1).

 {\small{
\begin{algorithm}[!tp]
\caption{\textbf{FL over clustering CSMA/CA network}}
\begin{enumerate}[     1:]
\item ~\textbf{N} participants can uplink update simultaneously
\item All participants have initial weights with ${\textbf{W}}(0)$

\item ~\textbf{for} each iteration $t=1,2,3,\ldots..R$\textbf{ do}
  
\item ~~\ ~\textbf{for} each device $i=1,2,3,\ldots..A$ \textbf{do}
\item ~~~~~~~~~ \textbf{for} $m=1,2,3,\ldots M$,$\text{ where }M=\frac{|\mathfrak{O}_{(i)}|}{B_{i}}$\textbf{
(Local training steps)}
\item ~~~~~~~~~~~ $\nabla f_{i}(\textbf{W}_{i}^{t})=\frac{1}{M}\sum_{m=1}^{M}\nabla l(h_{(\textbf{W}_{i}^{t})}(X_{i}^{m},y_{i}^{m}))$
\item ~~~~~~~~~~~ $\textbf{W}_{(i)}^{t}\longleftarrow \textbf{W}_{(i)}^{t}-\eta_{i}^{t}\nabla f_{i}(\textbf{W}_{i}^{t};B_{i})$
\item ~~~~~~~~~ \textbf{end}

\item ~~ FL Server receive ${\textbf{W}}_{{i}}^{t},{\nabla}f_{{i}}({\textbf{W}}_{{i}}^{t}),$ \textbf{from $A\leq K$ devices}
\item ~~~~~~~~~~~~~~ If $\frac{1}{A}(\sum_{{i}=1}^{A}{\nabla}f_{{i}}({\textbf{W}}_{{i}}^{t}))\thinspace\thinspace|A>1\thinspace]\thinspace\le\thinspace\varepsilon_{k}$
\item \textbf{~~~~~~~~~~~~~~~~ ~~~~Yes:} end process
(Gradient Convergence) 		
\item \textbf{~~~~~~~~~~~~~~~~~~ No:} continue
\item ~~~~~~~~~~~ $\hat{\textbf{W}}_{i}^{t}\longleftarrow(\sum_{{i}=1}^{A}(p_{i}{\textbf{W}}_{{i}}^{t}))$
\item ~~~~~~~~~~~ $\textbf{W}_{i}^{(t+1)}\longleftarrow\hat{\textbf{W}}_{i}^{t}$
\item ~~ FL Server broadcast $\hat{\textbf{W}}_{{i}}^{t},~ \text{go back to 3}$
\item ~~~\textbf{end}
\item ~\textbf{end}
\end{enumerate}
\end{algorithm}
}}

\section{Simulation results and discussion} \label{sec: Simulation and results}
\begin{figure}[t]
\centering
\includegraphics[width=3.0 in,height=2.2 in]{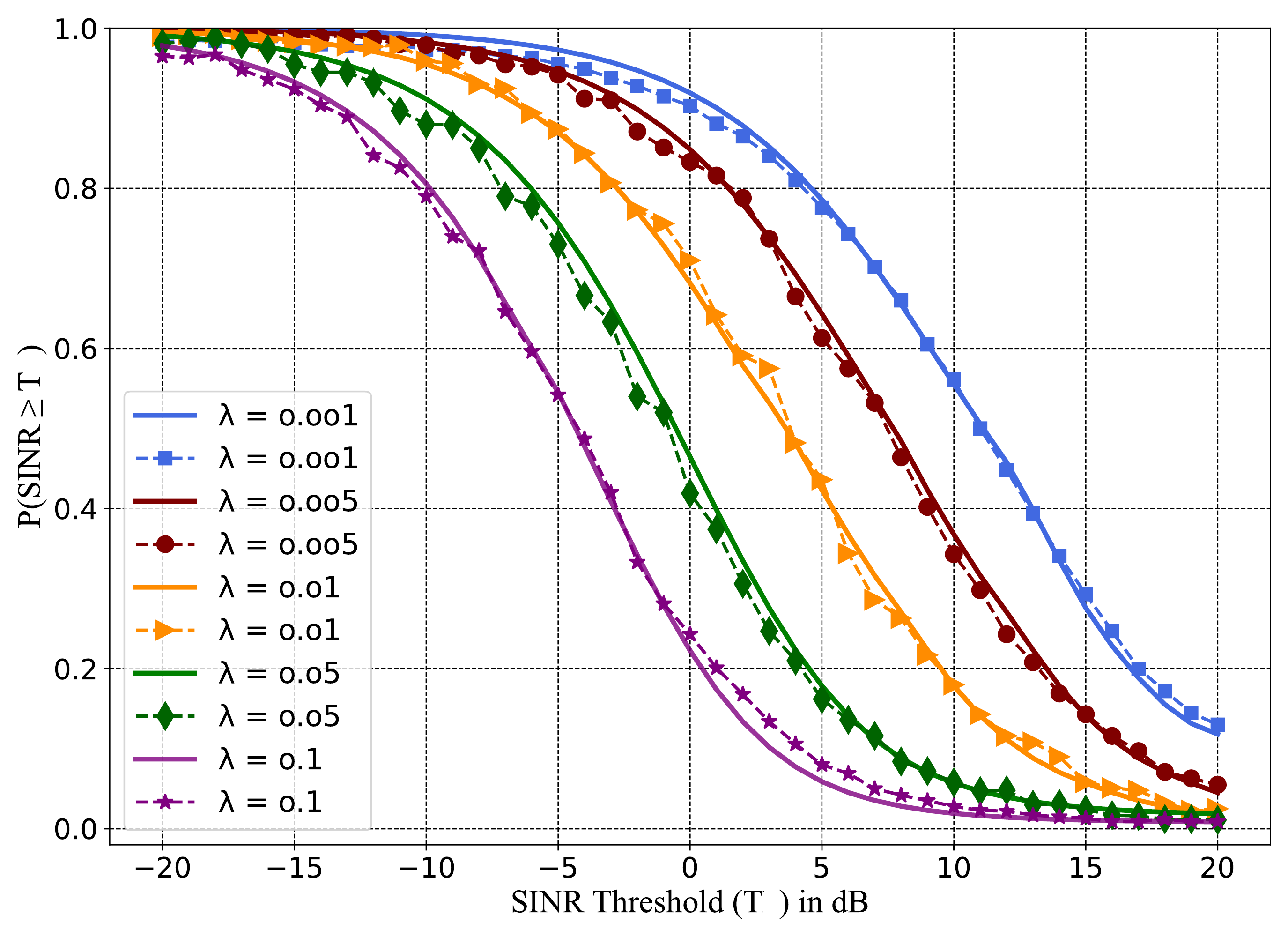}
\caption{The theoretical (solid lines) and simulation (dashed with markers) successful transmission probability as a function in $T$ in dB.}
\label{fig4}
\end{figure}

In the proposed network, a cellular frequency reuse technique (i.e., $N$ frequencies for the whole network) is used over a large-scale network of uniform cells in order to increase the capacity without increasing its allocated bandwidth during the learning process. The simulated network consists of random participants of both IoMT devices and untrusted devices (i.e, militia and hacker devices) that have a Poisson distribution within the target areas. First, the interference during the learning process in our large-scale network is evaluated in our simulation based on different values for $\lambda$ of the network. The results are shown in Figure \ref{fig4} for the proposed wireless communication model for different network intensities confirms the conclusions  drawn from the theoretical analysis in (\ref{eq: Max Cap CSMA}), where to achieve the desired capacity and accommodate the number of users in the FL network during the learning process, a trade-off between the probability of successful transmission and the SINR threshold is necessary. In comparison, the physical interference model and the performance model are evaluated where the same range of frequencies $N$ is used for the proposed network area, but the FL network in the baseline is presented without considering both (a) the untrusted devices that can diverge the learning process, and (b) the interference within the target area where the participants can use any of the available carrier frequencies randomly following the CSMA/CA protocol to uplink the model update. 

\begin{figure}
\centering
\includegraphics[width= 3.3 in, height=3.0 in]{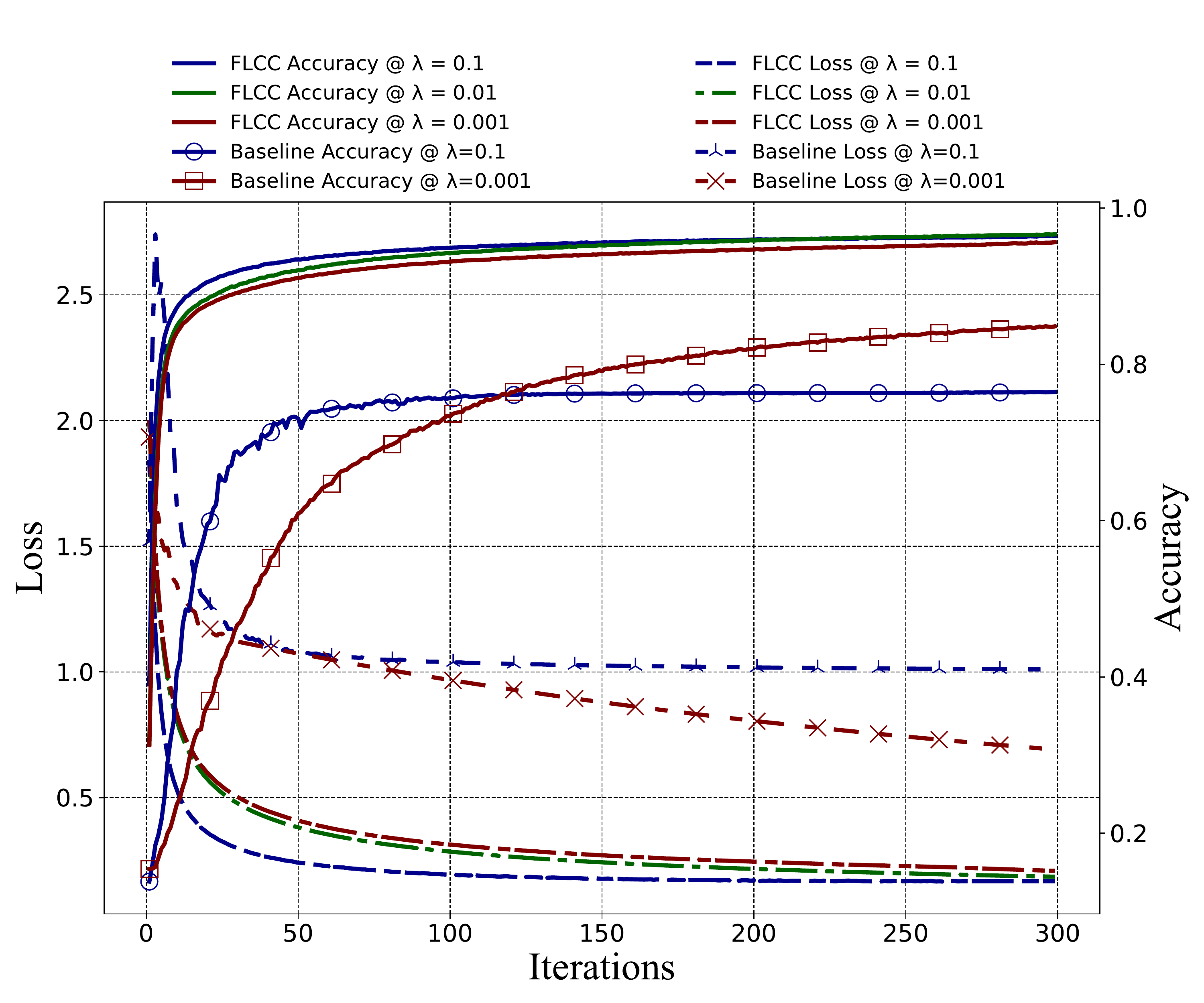}
\caption{Accuracy and loss for FLCC and baselines networks over CSMA/CA protocol.}
\label{fig:ACC_Loss}
\end{figure}

Similar to healthcare classification problems, our model is trained on the well-known MNIST dataset for benchmarking the proposed model, which contains 60K images samples of handwritten digits where each node within the network is assigned random samples in the range between (100-200) samples instance for training and validation using the Convolution Neural Network algorithm (CNN) \cite{xin2019research} as a local classified model. The performance metrics (i.e., the accuracy and loss) are evaluated for the proposed FLCC network and baselines at different network intensities for 300 iterations. Furthermore, the communication system model illustrates the relation between the SINR threshold ($T$) and the probability of successful transmission to define the number of successfully connected devices during the learning process. The FLCC model outcomes significantly improve performance in terms of accuracy and loss at variant $\lambda$ compared with the baselines model. In 300 iterations, the FLCC model achieved 98\% accuracy and less than 0.05 loss for different intensities. In contrast, the baselines (i.e., FL models over CSMA/CA but without eliminating and solving either the untrusted devices attack and the frequency interference issues) had a low performance  due to the untrusted devices and the interference in the network which are recorded a high loss and achieved only 75\% and 85\% model accuracy for 0.1 and 0.001 network intensities, respectively. 

As shown in Figure \ref{fig:ACC_Loss}, the improvement in FLCC model performance increases rapidly as a result of minimizing  the negative impact of less trusted and untrusted devices (i.e., hackers devices) as well as increases the probability of successful transmission of the IoMT nodes by reducing the interference via implementing the frequencies reuse technique for utilising the available frequencies channels by servers and devices within the network. 

\section{Conclusion} \label{sec: conclusion}
This study implements an FL model over CSMA/CA to preserve the privacy of IoMT users. However, the interference accrued in the learning process can affect negatively the performance of the FL model. In our work, we design a network in form of a clustering layout to optimize two major benefits for the network outcomes: (a) minimize collision; the nodes within the network exchange parameters with the server efficiently, (b) increase the performance; the model achieves high accuracy and low loss in a few numbers of iterations. Furthermore, combining the frequency reuse scheme with the clustered network results in a comprehensive improvement in the model performance, and the FLCC outcomes show a competitively low collision rate in comparison with traditional FL over CSMA/CA models. 

To sum up, simulating an efficient FL over CSMA/CA network and exchanging parameters during the distributed learning process should not only consider a fixed estimated number of users, which is traditionally used in simulation, but also the untrusted devices elimination, the user received signal strength, network intensity and communication protocols compatible with the existing networks.

In future work, a network segmentation technique can be integrated into the system and used to improve the overall FL network performance.

\nocite{*} 
\bibliographystyle{IEEEtran}
\bibliography{Abdelaziz_main_text}

\begin{thebibliography}{10}
\providecommand{\url}[1]{#1}
\csname url@samestyle\endcsname
\providecommand{\newblock}{\relax}
\providecommand{\bibinfo}[2]{#2}
\providecommand{\BIBentrySTDinterwordspacing}{\spaceskip=0pt\relax}
\providecommand{\BIBentryALTinterwordstretchfactor}{4}
\providecommand{\BIBentryALTinterwordspacing}{\spaceskip=\fontdimen2\font plus
\BIBentryALTinterwordstretchfactor\fontdimen3\font minus
  \fontdimen4\font\relax}
\providecommand{\BIBforeignlanguage}[2]{{%
\expandafter\ifx\csname l@#1\endcsname\relax
\typeout{** WARNING: IEEEtran.bst: No hyphenation pattern has been}%
\typeout{** loaded for the language `#1'. Using the pattern for}%
\typeout{** the default language instead.}%
\else
\language=\csname l@#1\endcsname
\fi
#2}}
\providecommand{\BIBdecl}{\relax}
\BIBdecl

\bibitem{li2013towards}
F.~Li, M.~V{\"o}gler, M.~Clae{\ss}ens, and S.~Dustdar, ``Towards automated iot
  application deployment by a cloud-based approach,'' in \emph{2013 IEEE 6th
  international conference on service-oriented computing and
  applications}.\hskip 1em plus 0.5em minus 0.4em\relax IEEE, 2013, pp. 61--68.

\bibitem{mcmahan2017communication}
B.~McMahan, E.~Moore, D.~Ramage, S.~Hampson, and B.~A. y~Arcas,
  ``Communication-efficient learning of deep networks from decentralized
  data,'' in \emph{Artificial intelligence and statistics}.\hskip 1em plus
  0.5em minus 0.4em\relax PMLR, 2017, pp. 1273--1282.

\bibitem{ieee2007part}
I.~.~W. Group \emph{et~al.}, ``Part 11: wireless lan medium access control
  (mac) and physical layer (phy) specifications,'' \emph{ANSI/IEEE Std 802.11},
  2007.

\bibitem{vishnu2020internet}
S.~Vishnu, S.~J. Ramson, and R.~Jegan, ``Internet of medical things (iomt)-an
  overview,'' in \emph{2020 5th international conference on devices, circuits
  and systems (ICDCS)}.\hskip 1em plus 0.5em minus 0.4em\relax IEEE, 2020, pp.
  101--104.

\bibitem{DIRANI2011141}
\BIBentryALTinterwordspacing
M.~Dirani, Z.~Altman, and M.~Salaun, ``Chapter 7 - autonomics in radio access
  networks,'' in \emph{Autonomic Network Management Principles}, N.~Agoulmine,
  Ed.\hskip 1em plus 0.5em minus 0.4em\relax Oxford: Academic Press, 2011, pp.
  141--166. [Online]. Available:
  \url{https://www.sciencedirect.com/science/article/pii/B9780123821904000073}
\BIBentrySTDinterwordspacing

\bibitem{haenggi2009stochastic}
M.~Haenggi, J.~G. Andrews, F.~Baccelli, O.~Dousse, and M.~Franceschetti,
  ``Stochastic geometry and random graphs for the analysis and design of
  wireless networks,'' \emph{IEEE journal on selected areas in communications},
  vol.~27, no.~7, pp. 1029--1046, 2009.

\bibitem{jing2012achievable}
T.~Jing, X.~Chen, Y.~Huo, and X.~Cheng, ``Achievable transmission capacity of
  cognitive mesh networks with different media access control,'' in \emph{2012
  Proceedings IEEE INFOCOM}.\hskip 1em plus 0.5em minus 0.4em\relax IEEE, 2012,
  pp. 1764--1772.

\bibitem{gupta2000capacity}
P.~Gupta and P.~R. Kumar, ``The capacity of wireless networks,'' \emph{IEEE
  Transactions on information theory}, vol.~46, no.~2, pp. 388--404, 2000.

\bibitem{ho2019real}
Y.~Ho and S.~Wookey, ``The real-world-weight cross-entropy loss function:
  Modeling the costs of mislabeling,'' \emph{IEEE Access}, vol.~8, pp.
  4806--4813, 2019.

\bibitem{sutton2018reinforcement}
R.~S. Sutton and A.~G. Barto, \emph{Reinforcement learning: An
  introduction}.\hskip 1em plus 0.5em minus 0.4em\relax MIT press, 2018.

\bibitem{xin2019research}
M.~Xin and Y.~Wang, ``Research on image classification model based on deep
  convolution neural network,'' \emph{EURASIP Journal on Image and Video
  Processing}, vol. 2019, no.~1, pp. 1--11, 2019.

\bibitem{bianchi2000performance}
G.~Bianchi, ``Performance analysis of the ieee 802.11 distributed coordination
  function,'' \emph{IEEE Journal on selected areas in communications}, vol.~18,
  no.~3, pp. 535--547, 2000.

\bibitem{castro2017survey}
D.~Castro, W.~Coral, J.~Cabra, J.~Colorado, D.~M{\'e}ndez, and L.~Trujillo,
  ``Survey on iot solutions applied to healthcare,'' \emph{Dyna}, vol.~84, no.
  203, pp. 192--200, 2017.

\bibitem{ZIOUVA2002313}
\BIBentryALTinterwordspacing
E.~Ziouva and T.~Antonakopoulos, ``Csma/ca performance under high traffic
  conditions: throughput and delay analysis,'' \emph{Computer Communications},
  vol.~25, no.~3, pp. 313--321, 2002. [Online]. Available:
  \url{https://www.sciencedirect.com/science/article/pii/S0140366401003693}
\BIBentrySTDinterwordspacing

\end{thebibliography}

\end{document}